\documentclass[conference]{IEEEtran}
\IEEEoverridecommandlockouts
\usepackage{comment}

\usepackage{cite}
\usepackage{amsmath,amssymb,amsfonts}
\usepackage{algorithmic}
\usepackage{graphicx}
\usepackage{textcomp}
\usepackage{tabularx}
\usepackage{hyperref}
\usepackage{setspace}
\usepackage{booktabs}
\usepackage{array}
\usepackage{tikz}
\usepackage[table]{xcolor}
\usepackage{mdframed}
\usepackage{caption}
\mdfdefinestyle{graybox}{
    backgroundcolor=gray!12,
    linecolor=gray!12,
    linewidth=0.0pt,
    roundcorner=3pt,
    skipabove=4pt,
    skipbelow=2pt,
    innertopmargin=4pt,
    innerbottommargin=4pt,
    innerleftmargin=4pt,
    innerrightmargin=4pt
}

\definecolor{codegreen}{rgb}{0,0.6,0}
\definecolor{codegray}{rgb}{0.5,0.5,0.5}
\definecolor{codepurple}{rgb}{0.58,0,0.82}
\definecolor{backcolour}{rgb}{0.95,0.95,0.92}

\usepackage[most]{tcolorbox}  

\tcolorboxenvironment{findings}{
    colback=blue!5,        
    colframe=blue!50!black,
    fonttitle=\bfseries,
    boxrule=0.5pt,         
    arc=2mm,               
    left=2mm, right=2mm, top=1mm, bottom=1mm, 
    before skip=2mm, after skip=2mm, 
    sharp corners=false,
    boxsep=1mm,
    size=small,
}

\definecolor{participantblue}{RGB}{80, 130, 200}   
\definecolor{lightbluebg}{RGB}{235, 245, 255}     

\newtcolorbox{participantquote}[1][]{
  enhanced,
  colback=lightbluebg,
  colframe=participantblue,
  boxrule=0.5pt,
  arc=2pt,
  left=8pt,
  right=8pt,
  top=5pt,
  bottom=5pt,
  before skip=6pt,
  after skip=6pt,
  borderline west={2pt}{0pt}{participantblue},
  #1
}
\usepackage{xcolor}
\def\BibTeX{{\rm B\kern-.05em{\sc i\kern-.025em b}\kern-.08em
    T\kern-.1667em\lower.7ex\hbox{E}\kern-.125emX}}

\usepackage{tikz}
\usepackage{xcolor}

\definecolor{TUMDarkGray}{RGB}{88,88,90}
\definecolor{TUMAccentLightBlue}{RGB}{152,198,234}
\definecolor{TUMAccentOrange}{RGB}{227,114,34}
\definecolor{TUMAccentGreen}{RGB}{162,173,0}
\definecolor{TUMLightGray}{RGB}{217,217,217}
\definecolor{TUMSecondaryBlue}{RGB}{0,82,147}

\begin{document}

\title{LegacyWorld: Atomicity-Aware Evaluation of GUI Agents for Legacy Workflows\\

}

\IEEEoverridecommandlockouts

\IEEEoverridecommandlockouts

\IEEEoverridecommandlockouts

\author{
\IEEEauthorblockN{
Thilo Reintjes\IEEEauthorrefmark{1}\textsuperscript{\S},
Sivajeet Chand\IEEEauthorrefmark{2}\textsuperscript{\S},
Derui Zhu\IEEEauthorrefmark{2},
Sushant Kumar Pandey\IEEEauthorrefmark{3},
Alexander Pretschner\IEEEauthorrefmark{2}
\thanks{\textsuperscript{\S}These authors contributed equally.}
}

\IEEEauthorblockA{
\IEEEauthorrefmark{1}Schub, Germany \quad
\IEEEauthorrefmark{2}Technical University of Munich, Germany \quad
\IEEEauthorrefmark{3}University of Groningen, The Netherlands
}

\IEEEauthorblockA{
\IEEEauthorrefmark{1}thilo@schub.tech \quad
\IEEEauthorrefmark{2}\{sivajeet.chand, derui.zhu, alexander.pretschner\}@tum.de \quad
\IEEEauthorrefmark{3}s.k.pandey@rug.nl
}
}
\maketitle

\begin{abstract}
Legacy and legacy-like enterprise systems often remain difficult to modernize because critical workflows expose limited programmable interfaces and still require manual GUI interaction. This paper reports a pre-deployment evaluation study motivated by the development of \emph{legacy-use}, an industry-oriented framework for automating such workflows with multimodal LLM agents. During framework development, domain experts helped identify stateful workflows where successful demos are not sufficient: a failed agent run may still leave persistent invalid changes in business or healthcare records. We therefore evaluate computer-use agents using atomicity: a run should either complete the intended workflow correctly or fail without unintended persistent side effects. We construct a domain-expert-informed benchmark of 28 Windows GUI workflows, each specified with an initial state, goal state, and task-specific validator. We compare expert-crafted prompts with prompts generated from screen recordings of expert golden-path executions. Across six hosted computer-use agents, our results show that useful completion, safe failure, and non-atomic side effects are distinct operational profiles. We conclude that workflow capture, state validators, and atomicity-aware acceptance tests should be first-class requirements for AI-based legacy workflow automation.
\end{abstract}


\section{Introduction}
\label{sec:introduction}

Many organizations still rely on legacy systems for operationally critical workflows, especially in healthcare, public administration, accounting, and enterprise back-office work~\cite{anthony2024enabling, ogunwole2023modernizing, irani2023impact}. 
These systems often expose limited programmable interfaces and must instead be operated through remote desktop sessions or graphical user interfaces (GUIs)~\cite{bisbal1999legacy, de2016methodology}. 
Because replacing them is costly, risky, and disruptive, many high-value workflows remain manual and labor-intensive~\cite{de2016methodology, enjamuri2025ai, irani2023impact}. Multimodal large language model (LLM) agents offer a promising automation layer for such systems. 
Unlike traditional robotic process automation, which is often brittle with respect to UI changes~\cite{vanderaalst2018rpa, syed2020rpa}, computer-use agents can observe screenshots, interpret instructions, and execute GUI actions at runtime~\cite{wang2025opencua, xie2024osworld, bonatti2024windows}. 
This makes them attractive for systems that are difficult to modernize from the inside: an agent can operate the same interface used by human workers.

This paper is motivated by \emph{legacy-use}, a framework for automating legacy GUI workflows with computer-use agents~\cite{legacyuse2026}. During development, knowledge workers emphasized that successful demos are insufficient for stateful enterprise workflows: deployment also depends on what state an agent leaves after failure. An incorrect click, partial form edit, or premature confirmation may durably modify operational state without completing the intended workflow, potentially corrupting records, violating process constraints, or requiring costly manual recovery in healthcare or administrative settings~\cite{bisbal1999legacy, de2016methodology, irani2023impact, iso25010}.

Existing computer-use benchmarks such as Windows Agent Arena~\cite{bonatti2024windows} and OSWorld~\cite{xie2024osworld} evaluate realistic desktop-task completion, but do not foreground whether failed or nominally successful attempts leave persistent state valid. WorkArena-style enterprise benchmarks~\cite{drouin2024workarena} cover business workflows, but typically through structured browser-level actions rather than raw GUI interaction with legacy-style environments. Classical RPA studies GUI automation mainly as scripted replay and its fragility under UI change. Our contribution is therefore atomicity-aware evaluation of probabilistic computer-use agents in stateful legacy workflows.

We construct a domain-expert-informed benchmark of 28 Windows GUI workflows across legacy and legacy-like enterprise applications. Each workflow is encoded as an executable task contract with an initial state, goal state, optional runtime parameters, and validators for intended effects and unintended side effects. Each fresh-VM run is evaluated by combining the agent-reported outcome with independently verified post-run state validity, yielding four classes: valid success, invalid success, valid failure, and invalid failure. A second practical challenge is workflow specification: manually writing detailed procedural prompts for every \emph{legacy-use} customer workflow does not scale. A longer-term direction is demonstration-driven automation, where a domain expert performs a workflow once, the system derives an agent-facing procedure from the recording, and the agent later executes it under the same task contract. We compare expert-crafted prompts with prompts generated from expert golden-path recordings. Videos supply only the procedural body; the initial state, goal state, parameters, response shape, and validators remain fixed. This is a controlled step toward learning procedures from demonstrations without treating recordings as correctness oracles.

Our study addresses two research questions:
   
    \textbf{RQ1:} \textit{How reliably do current computer-use agents execute domain-expert-informed legacy-style GUI workflows without leaving invalid persistent state?}

    \textbf{RQ2:} \textit{What changes in observed completion and state-validity outcomes when expert-crafted prompts are replaced by prompts generated from screen recordings?}

This paper makes three contributions. First, it distills from \emph{legacy-use} a pre-deployment lesson: successful GUI-agent demonstrations are insufficient for stateful legacy workflows unless failed executions are also evaluated. Second, it presents a domain-expert-informed benchmark of 28 GUI workflows, including 18 externally verified administrative, dental practice-management, and clinical tasks grounded through expert review and, where possible, production or staging checks. Third, it reports an atomicity-aware evaluation of computer-use agents across expert-crafted and video-generated specifications, showing that useful completion, safe failure, and non-atomic side effects must be measured separately.

\begin{figure}
\centering
\includegraphics[width=\columnwidth]{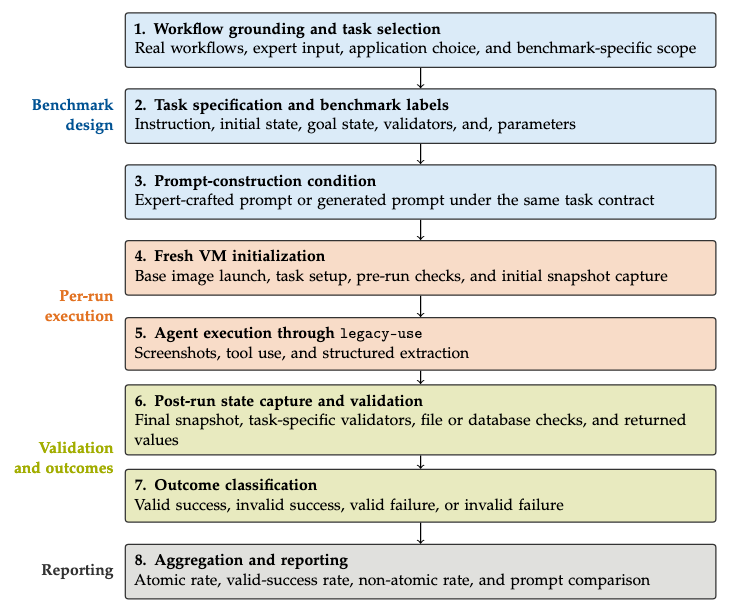}
\caption{Atomicity-aware evaluation workflow.}
\label{fig:evaluation-workflow}
\end{figure}

\section{Benchmark and Evaluation Method}
\label{sec:method}

\begin{table*}[t]
\centering
\caption{Task-level overview of the benchmark. SC = state-changing intended workflow; EV = externally verified by experts.}
\label{tab:task-level-overview}
\footnotesize
\renewcommand{\arraystretch}{1.04}
\setlength{\tabcolsep}{3pt}
\begin{tabular*}{\textwidth}{@{\extracolsep{\fill}}lllccl}
\toprule
\textbf{App.} & \textbf{Task} & \textbf{Domain} & \textbf{SC} & \textbf{EV} & \textbf{Primary validator} \\
\midrule
Adempiere & Create Dunning Notice & Admin/ERP & Yes & Yes & File-change check \\
Adempiere & Open Adempiere & Admin/ERP & No & Yes & Returned user ID \\
Adempiere & Setup Payroll & Admin/ERP & Yes & Yes & Generated business values \\
Calculator & Add two numbers & Reference & No & No & Returned result \\
Chrome & Install Chrome & Reference & Yes & No & Executable exists \\
Crystal & Create PDF & Reporting & Yes & Yes & Exported PDF exists \\
Crystal & Load Report & Reporting & No & Yes & Returned report title \\
Crystal & Open Viewer & Reporting & No & Yes & Window-state check \\
DSWin & Add Billing Code & Dental/healthcare & Yes & Yes & Database delta \\
DSWin & Add New Patient & Dental/healthcare & Yes & Yes & Database delta \\
DSWin & Add Patient Documentation & Dental/healthcare & Yes & Yes & Database delta \\
DSWin & Look Up Patient & Dental/healthcare & No & Yes & Returned patient data \\
DSWin & Open DSWin & Dental/healthcare & No & Yes & Manifest/window check \\
DSWin & Update Patient Phone & Dental/healthcare & Yes & Yes & Database delta \\
IrfanView & Batch Convert Images & Utility & Yes & No & Output-file check \\
IrfanView & Resize Image & Utility & Yes & No & Output hash \\
Notepad & Save File & Reference & Yes & No & File content \\
Notepad & Resume \& Save & Reference & Yes & No & File content \\
Open Clinic & Add Operation Report & Clinical & Yes & Yes & Returned operation slot \\
Open Clinic & Collect Patient Data & Clinical & No & Yes & Returned patient data \\
Open Clinic & Look Up Patient & Clinical & No & Yes & Returned record number \\
Open Clinic & Onboard Patient & Clinical & Yes & Yes & Returned record number \\
Open Clinic & Open Application & Clinical & No & Yes & Returned URL \\
OpenMRS & Onboard Patient & Clinical & Yes & Yes & Patient system data \\
VS Code & Add Python Extension & Developer tool & Yes & No & Extension state \\
VS Code & Look Up Extension & Developer tool & No & No & Manifest unchanged \\
VS Code & Open VS Code & Developer tool & No & No & Window/workspace state \\
Windows & Hello World & Reference & Yes & No & File content \\
\bottomrule
\end{tabular*}
\end{table*}

This work originates from \textit{legacy-use}\cite{legacyuse2026}, a framework for automating legacy GUI workflows with computer-use agents. Domain-expert discussions in healthcare, administration, and enterprise settings showed that task success alone is insufficient: failed executions may still leave persistent side effects in files, records, or application state, motivating our atomicity-aware evaluation design. Figure~\ref{fig:evaluation-workflow} summarizes the benchmark lifecycle: workflow grounding, task-contract construction, prompt-condition selection, fresh-VM execution, post-run state capture, validation, four-way outcome classification, and aggregate reporting. Table~\ref{tab:task-level-overview} compactly lists tasks by application, domain, state-changing behavior, external verification, and primary validator type. We distinguish state-changing tasks because unsafe side effects are most consequential when workflows modify persistent files, records, application configuration, installed software, or exported artifacts. 

\subsection{Workflow Selection and Task Contracts}
\label{subsec:workflow-selection}

The task suite balances practical relevance with controlled evaluation. Eighteen tasks were externally verified by domain experts or operational staff as representative of real-world work. This verification refers to design-time review of workflow realism and task specification, not independent auditing of model outputs. Experts from healthcare, administrative, med-tech startup, university-hospital, and enterprise contexts helped identify manual workflows, review action sequences, and confirm that modeled goal states matched operational expectations. For externally grounded tasks, workflows were checked during benchmark design on production systems or, where infeasible, staging systems; all reported executions were then run in isolated virtual machines. Thus, external verification grounds the tasks and contracts, but does not imply production deployment, production evaluation, or independent certification.

Each workflow is converted into an executable task contract containing a natural-language instruction, explicit initial-state assumptions, a goal-state definition, runtime parameters where needed, an expected return-value schema for tasks requiring extracted identifiers or values, and task-specific validation logic. The contract also records whether the workflow is state-changing or externally grounded, and which benchmark-relevant state changes are allowed or forbidden. This is necessary because, for stateful GUI automation, a prompt alone does not define acceptable execution: the benchmark must specify intended changes, forbidden side effects, and how the final state is checked.

The tasks vary in the type of state they modify or inspect. 
Some tasks are non-destructive lookup or setup tasks, such as opening an application or returning a value discovered during execution. 
Others modify persistent state, such as adding a new patient in DSWin, updating a patient phone number, creating a payroll setup in Adempiere, exporting a Crystal Reports PDF, or onboarding a patient in Open Clinic or OpenMRS. 
This mix reflects the deployment concern that motivated the benchmark: in legacy-style workflows, failed attempts may still leave durable state changes behind.

\subsection{Validation Logic}
\label{subsec:validation}

Validation combines generic snapshot checks with task-specific validators and is independent of the agent's own success report. Generic checks compare pre-run and post-run benchmark state, for example required or forbidden files, directory manifests, hashes, or window state. Task-specific validators inspect application-dependent artifacts such as database deltas, exported reports, returned identifiers, URLs, or calculated values. For example, the Windows Hello World task checks that the expected file exists with exact contents; DSWin patient-record tasks compare pre-run and post-run database snapshots and accept only the intended row or field changes; and lookup tasks validate returned identifiers or values against task-specific ground truth.

Validators check intended effects, unintended side effects, and returned values. For state-changing tasks, state validity requires the intended change and no unexpected changes to monitored files, database rows, or application state. Failed runs are valid under task-specific conditions, typically unchanged monitored state. If an agent reports success but returns a wrong patient identifier, URL, slot, or calculated value, the run is an invalid success.

\subsection{Prompt Construction}
\label{subsec:prompt-construction}

We compare two ways of specifying workflows to agents. 
The first condition uses expert-crafted prompts. 
These prompts encode the intended workflow as step-by-step procedural instructions, informed where applicable by domain experts and knowledge workers familiar with the corresponding software.

The second condition evaluates demonstration-derived workflow specification in a controlled form. For each task, \textit{gemini-3.1-pro-preview}\cite{google2026gemini31propreview} generates one prompt from a single golden-path screen recording, a fixed English meta-prompt, and the canonical task interface from the expert-crafted condition, including task parameters and an expected response example. Thus, the generator produces only the procedural body of the task instruction, not the complete task contract. The generation output is constrained to structured action steps, which are rendered deterministically into the benchmark prompt format. 
No post-generation manual editing or selective cleanup is applied. 
The generated-prompt condition tests whether a procedure reconstructed from an expert demonstration can approximate an expert-written procedure under the same task contract. Recordings provide only pixels and temporal ordering, with highlighted mouse clicks but no separate keyboard or mouse event log; focus changes, shortcuts, off-screen typing, transient feedback, fast interactions, or implicit domain assumptions may be weakly visible or absent. We therefore treat recordings as workflow-capture aids, not correctness oracles: the initial state, goal state, parameters, response shape, and validators remain fixed. Thus, video generation reconstructs a procedure under a fixed task contract rather than learning the full task from video; the recording specifies how the workflow is attempted, while the contract specifies what state is correct.

\subsection{Execution Environment}
\label{subsec:execution}

Each run is executed in a fresh Windows virtual machine derived from a fixed base image. The benchmark performs task setup, records a pre-run snapshot of benchmark-relevant state, launches the agent, and after termination, timeout, or \texttt{ui\_not\_as\_expected}, captures a final snapshot and applies the task validators. State changes are stored in a temporary overlay that is discarded after validation, ensuring comparable initial state across runs.

Agents interact through the \emph{legacy-use} harness~\cite{legacyuse2026}, which provides screenshot-based observation, GUI actions, structured extraction, and an explicit safe-stop path. All agents used the same benchmark-facing configuration: fresh VM per run, 900-second timeout, \texttt{max\_tokens=4096}, one included run per task and prompt condition, and the same task contract with either expert-crafted or video-generated prompts. Provider-side API or infrastructure faults were rerun or excluded before aggregation; model-visible non-success terminations were retained as agent failures and evaluated for post-run state validity. We evaluated six hosted computer-use agents through this shared harness: OpenAI GPT-5.4 (gpt-5.4)\cite{openai_gpt54_2026}, Google Gemini 2.5 Computer Use (gemini-2.5-computer-use-preview-10-2025)\cite{google_gemini25_computer_use_2026}, Anthropic Claude Opus 4.6 (anthropic.claude-opus-4-6-v1)\cite{anthropic_claude_models_2026}, Anthropic Claude Sonnet 4.6 (global.anthropic.claude-sonnet-4-6)\cite{anthropic_claude_models_2026}, Anthropic Claude Haiku 4.5 (global.anthropic.claude-haiku-4-5-20251001-v1:0)\cite{anthropic_claude_models_2026}, and Moonshot AI Kimi K2.5 (moonshotai.kimi-k2.5)\cite{moonshot_kimi_k25_2026}. Across agents, the benchmark-facing call used \texttt{temperature=0.0}; provider defaults applied where parameters were not overridden.

\begin{table*}[t]
\centering
\caption{Operational reliability dashboard for expert-crafted and video-generated prompts.}
\label{tab:aggregate-results}
\footnotesize
\renewcommand{\arraystretch}{1.08}

\begin{tabular*}{\textwidth}{@{\extracolsep{\fill}}lrrrrrrrrrr}
\toprule
\textbf{Agent} &
\multicolumn{5}{c}{\textbf{Expert-crafted prompts}} &
\multicolumn{5}{c}{\textbf{Video-generated prompts}} \\
\cmidrule(lr){2-6} \cmidrule(lr){7-11}
& \textbf{VS} & \textbf{VF} & \textbf{IS} & \textbf{IF} & \textbf{US}
& \textbf{VS} & \textbf{VF} & \textbf{IS} & \textbf{IF} & \textbf{US} \\
\midrule
\texttt{gpt-5-4}\cite{openai_gpt54_2026}    &  3.6 & 96.4 &  0.0 & 0.0 &  0.0 &  3.6 & 96.4 &  0.0 & 0.0 &  0.0 \\
\texttt{gemini-2.5}\cite{google_gemini25_computer_use_2026} &  7.1 & 85.7 &  0.0 & 7.1 &  7.1 & 10.7 & 89.3 &  0.0 & 0.0 &  0.0 \\
\texttt{opus-4-6}\cite{anthropic_claude_models_2026}   & 78.6 & 10.7 & 10.7 & 0.0 & 10.7 & 50.0 & 39.3 &  7.1 & 3.6 & 10.7 \\
\texttt{haiku-4-5}\cite{anthropic_claude_models_2026}  & 71.4 & 14.3 & 14.3 & 0.0 & 14.3 & 46.4 & 39.3 & 14.3 & 0.0 & 14.3 \\
\texttt{sonnet-4-6}\cite{anthropic_claude_models_2026} & 75.0 &  7.1 & 14.3 & 3.6 & 17.9 & 53.6 & 28.6 & 14.3 & 3.6 & 17.9 \\
\texttt{kimi-k2.5}\cite{moonshot_kimi_k25_2026}  & 42.9 & 21.4 & 28.6 & 7.1 & 35.7 & 25.0 & 39.3 & 32.1 & 3.6 & 35.7 \\
\bottomrule
\end{tabular*}

\vspace{2pt}
\begin{minipage}{0.98\textwidth}
\footnotesize
\emph{Note.} VS = valid success, VF = valid failure, IS = invalid success, IF = invalid failure, 
US = unsafe side-effect rate (IS+IF). Values are percentages over 28 tasks per condition. Atomicity is VS+VF, equivalently 100-US.
\end{minipage}
\end{table*}

\subsection{Atomicity-Aware Outcome Model}
\label{subsec:outcome-model}

Each run is evaluated along two dimensions: the agent-reported terminal status and independently verified post-run state validity. Their combination yields four outcome classes: \textbf{valid success}, where the agent reports completion and reaches the goal state; \textbf{invalid success}, where it reports completion but leaves invalid or unintended changes; \textbf{valid failure}, where it fails, stops, or times out while preserving the benchmark-relevant state; and \textbf{invalid failure}, where it does not complete the task and leaves unintended persistent changes.

We define \textbf{atomicity} as the fraction of runs ending in valid success or valid failure. It measures whether the system was left in an acceptable state, whereas valid success measures useful task completion. This distinction matters for deployment: a model may be highly atomic by failing conservatively yet complete few workflows, or complete many workflows while still causing non-atomic side effects unacceptable in stateful enterprise systems.

\subsection{Replication package}
The replication package is publicly available\footnote{\url{[https://github.com/ThiloReintjes/LegacyWorld}}. It contains the benchmark catalog, prompts, expert review, and validator metadata, run records, and scripts to reproduce the reported tables. For each of the 28 tasks, it documents the task contract, including initial and goal states, runtime parameters, return schema, permitted and forbidden side effects, and benchmark classifications. It also includes reviewer details, validators, oracle values, state artifacts, run outcomes, and runtime metadata.

\section{Results}
\label{sec:results}

We report the four-way outcome distribution rather than a single success metric because stateful GUI automation has two distinct operational concerns. First, an agent should complete useful work when possible. Second, regardless of whether it completes the task, it should leave the monitored system state acceptable. We therefore use valid success to describe useful completion and atomicity to describe state acceptability after execution. Atomicity includes both valid successes and valid failures: a safely failed run may be operationally preferable to a run that reports success while leaving incorrect persistent state. Each percentage reflects one included trajectory per task, model, and prompt condition. The results should therefore be read as descriptive evidence about operational profiles under this benchmark configuration, not as inferential rankings of model capability.

\subsection{RQ1: Atomic Execution of Legacy-Style Workflows}
\label{subsec:rq1-results}

Table~\ref{tab:aggregate-results} reports the full four-way outcome distribution for both prompt conditions. The expert-crafted condition illustrates why valid success and atomicity must be reported together. Some trajectories preserve state but rarely complete the workflow. For example, gpt-5-4 reaches 100.0\% atomicity in this condition, but almost all of this comes from valid failures: only 3.6\% of runs are valid successes, while 96.4\% are failures that preserve the monitored benchmark state. This is a conservative execution profile: it avoids invalid persistent changes, but provides little useful automation. Other trajectories show the complementary trade-off. For example, opus-4-6, sonnet-4-6, and haiku-4-5 produce substantially more valid successes under expert-crafted prompts, with valid-success rates of 78.6\%, 75.0\%, and 71.4\%, respectively. However, these useful completions are accompanied by non-atomic outcomes in some runs. Thus, higher useful completion does not imply that all executions leave the system in an acceptable state. The kimi-k2.5 results further illustrate why valid success alone is insufficient. Its expert-crafted condition contains a moderate valid-success rate of 42.9\%, but also the largest unsafe side-effect rate in this condition, 35.7\%. A deployment decision based only on task completion would miss this state-safety concern.

A single metric would obscure these profiles. Valid success measures correct workflow completion but not state damage in failed or misreported runs; atomicity measures an acceptable final state but can reward conservative non-completion. For stateful legacy-style workflows, valid success captures automation value, while atomicity captures state safety, so both must be reported with the four-way outcome distribution.

The invalid outcomes in Table~\ref{tab:aggregate-results} are not a single failure type. 
Across tasks, the validators expose several operationally different unsafe side effects: unexpected database deltas in patient-record workflows, unexpected file-system changes in export or file-creation workflows, invalid returned identifiers or calculated values, incorrect application or window state after setup tasks, and unsafe failures where the agent terminates after partial progress. 
One observed DSWin ``Add New Patient'' run illustrates why completion and state validity must be considered separately. The agent returned \texttt{status: success} and a structured patient record with plausible demographic and insurance fields. However, the post-run database validator rejected the run because the persisted \texttt{PATIENT.DBF} row violated the task contract: one insurance number was created \texttt{4212505} instead of the expected \texttt{1234567890}. This run is therefore an invalid success: it completed from the agent's perspective, but left durable patient-record state invalid. In contrast, a failed run that leaves the monitored database unchanged would be a valid failure and may be operationally preferable, because it requires re-execution rather than repair of incorrect persistent state.

This distinction matters operationally. A non-atomic outcome may correspond to a persistent record change, a wrong or extra file artifact, an incorrect discovered identifier, or an abandoned application state. In patient-record workflows, such cases would require human inspection or repair before the automated workflow could be trusted. Therefore, validators must check not only whether the agent reports completion, but whether the resulting persistent state is acceptable.

\begin{mdframed}[style=graybox]

\textbf{Summary RQ1.}
Current agents remain unreliable on expert-informed legacy GUI workflows, highlighting the need to evaluate both valid task success and atomicity to balance automation value with state safety.
\end{mdframed}

\subsection{RQ2: Expert-Crafted versus Video-Generated Prompts}
\label{subsec:rq2-results}

RQ2 evaluates whether expert screen recordings can support a more scalable workflow-specification process. 
The right half of Table~\ref{tab:aggregate-results} reports the video-generated prompt condition, where the procedural instructions are generated from golden-path recordings while the task contracts and validators remain unchanged. This evaluates a controlled step toward demonstration-driven workflow specification: recordings are used to recover the procedure, but the benchmark still supplies the task contract and determines correctness through validators. 
It does not test end-to-end task inference from demonstrations alone.

In the video-generated condition, several agents shift from valid success toward valid failure. For example, opus-4-6 changes from 78.6\% to 50.0\% valid success, sonnet-4-6 from 75.0\% to 53.6\%, haiku-4-5 from 71.4\% to 46.4\%, and kimi-k2.5 from 42.9\% to 25.0\%. In these cases, the main effect of video-generated prompts is not simply an increase in unsafe side effects; rather, useful completion decreases while many additional runs remain state-valid failures. By contrast, gpt-5-4 remains unchanged at 3.6\% valid success, and gemini-2.5 improves slightly from 7.1\% to 10.7\%.

The four-way table suggests a more specific interpretation than “video-generated prompts are less safe.” For several agents, lower valid success is accompanied by more valid failures rather than a proportional increase in invalid outcomes. For example, opus-4-6 keeps 89.3\% atomicity while valid success decreases from 78.6\% to 50.0\%, and haiku-4-5 keeps 85.7\% atomicity across both prompt conditions. In these cases, the generated procedure appears to reduce useful completion more than state acceptability.

These results reinforce the need to separate completion from state validity. Video-generated prompts can preserve atomicity while reducing valid success, so they should be evaluated by both endpoint completion and whether failed or partial attempts leave the monitored state acceptable.

\begin{mdframed}[style=graybox]

\textbf{Summary RQ2.}
Demonstration-based prompt generation can preserve state safety in some cases, but it currently reduces practical automation value compared with expert-crafted prompts.
\end{mdframed}

\section{Implications and Limitations}
\label{sec:implications-limitations}

The main practical lesson is that successful demos are not enough for stateful GUI automation. Teams should evaluate whether failed executions preserve an acceptable state, not only whether some runs complete. In our benchmark, this required explicit task contracts with an initial state, a goal state, allowed changes, forbidden side effects, and validators. Screen recordings are useful workflow-capture aids, but correctness still depends on post-run state rather than resemblance to a demonstrated trajectory.

This study remains a controlled pre-deployment evaluation. All runs were executed in isolated virtual machines, atomicity is defined only with respect to monitored observables, and each model--task--prompt cell contributes one included trajectory. The benchmark therefore supports risk-aware comparison of workflow automation profiles, not deployment certification or inferential ranking over stable stochastic distributions. The prompt comparison likewise studies reconstruction from a single golden-path recording under a fixed task contract rather than unconstrained task inference from demonstrations.

\section{Conclusion}
\label{sec:conclusion}

AI-based GUI automation is a plausible modernization layer for legacy systems, but task completion alone is not a sufficient deployment criterion. On 28 stateful Windows GUI workflows, we find that useful completion, safe failure, and non-atomic side effects are distinct operational profiles. We also find that prompts generated from expert screen recordings can support workflow capture, but do not replace explicit validation. For practice, the implication is simple: successful demos should be supplemented with atomicity-aware acceptance tests that check not only whether an agent completes a workflow, but also whether failed executions leave the system in an acceptable state.

\bibliographystyle{IEEEtran}
\bibliography{biblo}
\end{document}